\documentclass[sigconf,nonacm]{acmart}

\usepackage{graphicx}
\usepackage{booktabs}
\usepackage{arydshln}
\usepackage{balance}
\usepackage{multirow}

\AtBeginDocument{%
  \providecommand\BibTeX{{%
    \normalfont B\kern-0.5em{\scshape i\kern-0.25em b}\kern-0.8em\TeX}}}

\begin{document}

\title{Benchmarking Middle-Trained Language Models for Neural Search}

\author{Hervé Déjean}
\email{herve.dejean@naverlabs.com}
\orcid{0000-0002-9837-5358}
\affiliation{%
  \institution{Naver Labs Europe}
  \streetaddress{6 chemin de Maupertuis}
  \city{Meylan}
    \country{France}
}

\author{Stéphane Clinchant}
\email{stephane.clinchant@naverlabs.com}
\orcid{0000-0003-2367-8837}
\affiliation{%
  \institution{Naver Labs Europe}
  \streetaddress{6 chemin de Maupertuis}
  \city{Meylan}
    \country{France}
}
\author{Carlos Lassance}
\email{carlos.lassance@naverlabs.com}
\orcid{0000-0002-7754-6656}
\affiliation{%
  \institution{Naver Labs Europe}
  \city{Meylan}
  \country{France}
}

\author{Simon Lupart}
\email{simon.lupart@naverlabs.com}
\orcid{1234-5678-9012}
\affiliation{%
  \institution{Naver Labs Europe}
  \city{Meylan}
  \country{France}
}

\author{Thibault Formal}
\email{thibault.formal@naverlabs.com}
\orcid{0009-0008-6363-9553}
\affiliation{%
  \institution{Naver Labs Europe}
  \city{Meylan}
  \country{France}
}

\renewcommand{\shortauthors}{Hervé Déjean, Stéphane Clinchant, Carlos Lassance, Simon Lupart, \& Thibault Formal}

\begin{abstract}
Middle training methods aim to bridge the gap between the Masked Language Model (MLM) pre-training and the final finetuning for retrieval. Recent models such as CoCondenser, RetroMAE, and LexMAE argue that the MLM task is not sufficient enough to pre-train a transformer network for retrieval and hence propose various tasks to do so. Intrigued by those novel methods, we noticed that all these models used different finetuning protocols, making it hard to assess the benefits of middle training. We propose in this paper a benchmark of CoCondenser, RetroMAE, and LexMAE, under the same finetuning conditions. We compare both dense and sparse approaches  under various finetuning protocols and middle training on different collections (MS MARCO, Wikipedia or Tripclick). 
We use additional middle training baselines, such as a standard MLM finetuning on the retrieval collection, optionally augmented by a CLS predicting the passage term frequency.
For the sparse  approach, our study reveals that there is almost no statistical difference between those methods: the more effective the finetuning procedure is, the less difference there is between those models. 
For the dense approach, RetroMAE using MS MARCO as middle-training collection shows excellent results in almost all the settings.  
Finally, we show that middle training on the retrieval collection, thus adapting the language model to it, is a critical factor.
Overall, a better experimental setup should be adopted to evaluate middle training methods.
Code available at \url{https://github.com/naver/splade/tree/benchmarch-SIGIR23}


\end{abstract}




\maketitle

\section{introduction}
In recent years, Pre-trained Language Models (PLM) approaches, such as BERT \cite{devlin-etal-2019-bert}, have become dominant in the field of Information Retrieval \cite{passagereranking_bert_19,tonelloto_lecture_nir,yates-etal-2021-pretrained}. The common approach, as in many other domains, consists in finetuning a generic Pre-trained Language Model (hereafter LM) based on the Transformer architecture\cite{vaswani2017attention}. This applies to both the first stage and reranking steps, as well as various models -- including dense and sparse approaches \cite{Guo_2022}. 
However, it has been recently argued that the pre-training tasks for these models are not completely suited for IR. Hence, several works have introduced a middle training step to boost the final effectiveness \cite{gao-callan-2021-condenser,shen_lexmae_2022,liu_retromae_2022}. In practice, middle training is performed through additional unsupervised training steps aiming at preconditioning the network for the retrieval task. 
\begin{figure}
     \includegraphics[width=0.45\textwidth]{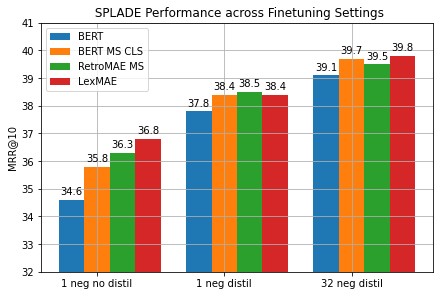}
    \caption{The more effective the finetuning procedure is, the less difference there is between SPLADE middle trained models.}
    \label{fig:splade}
\end{figure}
Our motivation in this paper is to better understand and assess the benefits of middle training. We notice that a fair comparison between these Middle-Trained Language Models (hereafter MTLMs) is missing in the literature, and drawing conclusions about their impacts is, therefore, challenging. Overall, the existing works are not directly comparable to each other as finetuning is composed of multiple steps and relies on the extraction of hard negatives to boost performance. In the end, it is unclear whether middle-trained models work well \textit{out of the box}, and which middle training is the most effective. 
This is why we study the impact of several of the recently proposed MTLMs, by comparing them under the same setting and testing several finetuning settings.
We thus conduct experiments considering well-established options, which will be further discussed in the methodology part.

To some extent, this benchmark is related to recent works that aim to understand the role of pre-training for language models \cite{tay-etal-2021-pretrained,Tay2022ScaleEI}. 
\citeauthor{pretraining_nlp_arxiv}~\cite{pretraining_nlp_arxiv} experiment with training LM from scratch on the downstream collections. More recently, \citeauthor{lassance_experimental_2023}~\cite{pretraining_scratch_arxiv23} examine the same questions in the IR context. All these recent works have shown some benefits thanks to pre-training from scratch. 
However, in this work, we focus on the recent middle training methods which showed additional gains and which did not only include a language model finetuning but also proposed other pre-training tasks. 
Contrary to the aforementioned works, our study wants to assess the role of those new pre-training tasks compared to the sole adaptation of the language model. 

Overall, this paper makes the following contributions: 
\begin{itemize}
    \item We benchmark several MTLMs using the MS MARCO and BEIR collections under the same conditions, by starting from a 'low-level' setting (1 negative, no distillation) to more effective 'high-level' settings (several negatives with distillation).
    \item For the sparse approach, we show that MTLMs bring some value for low-level settings (no distillation, less finetuning data), and have less impact with high-level settings.
    \item For the dense approach, RetroMAE performs the best in many settings, but it requires the use of the downstream collection during middle-training.
     \item Related to the last point, we compare MTLMs to the simple MLM finetuning on the retrieval collection and show that MTLMs are really the most effective when the retrieval collection is used for middle training. Hence, an important boost of MTLMs is implicitly due to the LM adaptation.
     In other words, their performance on zero-shot benchmarks such as BEIR, or with distribution shifts in another domain is not much better than the baseline.
\end{itemize}



\section{Related Work and Methodology}

\subsection{Middle Training Methods}
One of the earliest work on pre-training for retrieval has been explored in ~\cite{Chang2020Pre-training_ICLR}, where Wikipedia is used to perform an Inverse Cloze Task combined  with a  hyperlink prediction -- bringing benefits for retrieval tasks. Other forms of pre-training use a form of self-supervision with hyperlinks ~\cite{Ma2021Pretraining_hyperlink} or by doing a regression on the \textit{tfidf} scores of a document \cite{losses_pretraining_2020}.
In contrast, PROP~\cite{ma2020prop,ma2020bprop} propose pre-training with representative word predictions as a way to generate pseudo-relevant queries  for each document.


More recently, Condenser \cite{gao-callan-2021-condenser} revisits the idea of middle training and puts forward again the argument that the dense pooling mechanism is not well preconditioned for retrieval tasks. 
Hence, they proposed an additional task where the network has to rely on the CLS representation to perform an MLM prediction. An idea which has been named as \textit{information bottleneck} in later works ~\cite{kim-etal-2021-self}. 
Note that this is slightly different from other self-supervision tasks with pseudo-labeled data. Later, the Condenser architecture was refined in the coCondenser model 
\cite{gao-callan-2022-unsupervised} by adding a similarity task between different text spans from the same document.
coCondenser follows a two-round retriever training pipeline, inspired by the RocketQA paper \cite{gao-callan-2021-condenser,gao-callan-2022-unsupervised}. A first retriever is learned from the BM25 negatives, and then a second set of negatives is extracted. No details are given in the paper about the selection of those negatives (i.e., which top-$k$ those negatives are taken from and how many negatives are taken for each query). Surprisingly, coCondenser does \textit{not} use cross-encoder distillation \cite{hofstaetter2020_crossarchitecture_kd}, a proven technique to boost performance. In this paper, we use the coCondenser model being trained on MS MARCO\footnote{\url{Luyu/co-condenser-marco}}.

RetroMAE \cite{liu_retromae_2022}  develops further the idea of creating an information bottleneck, by masking twice an input passage so that the first masking produces a CLS representation reused to  decode the second masking of the passage. Furthermore, they consider a different way to perform their decoding, by explicitly duplicating the CLS representation for multiple masked predictions-- called enhanced decoding. In this paper, we use two available models: RetroMAE trained on WikiBook, the second, RetroMAE\_MSMARCO trained on MS MARCO\footnote{on HuggingFace: Shitao/RetroMAE and Shitao/RetroMAE\_MSMARCO}.

Very recently, LexMAE\cite{shen_lexmae_2022} has achieved state-of-the-art results by adopting a bottleneck pre-training for the SPLADE model \cite{formal_splade_2021}. As in RetroMAE, an input passage is masked twice, and a bottleneck representation is created by a weighted average of embedding of terms predicted by the encoder. Then, this pooled representation is used as a way to condense the information of the first passage, which is then reused to perform MLM on the second passage.  The middle-trained model we use in this paper is trained on MS MARCO, and was provided by the authors.
During finetuning, LexMAE employed a 3-step finetuning pipeline, with BM25 negatives, its own negatives, and finally, a distillation step from a cross-encoder.

All in all, those recent middle training methods employ different finetuning protocols, with different negatives or cross-encoder teachers, making a fair comparison of the models hard. This is why we propose to test those pre-trained methods in the same \textit{conditions}: a) with and without distillation with a fixed set of negatives and b) either with 1 or more hard negatives. Distillation is, in fact, an important source of performance gain used in the majority of works \cite{lin-etal-2021-batch, santhanam2021colbertv2,Hofstaetter2021_tasb_dense_retrieval,splade_sigir22, lassance2022efficiency}. We train our models with hard negatives from multiple sources and distillation scores from the best available cross-encoder\footnote{\url{https://huggingface.co/cross-encoder/ms-marco-MiniLM-L-6-v2}}, to our knowledge, with a 
 KL-Div loss.\footnote{\url{https://huggingface.co/datasets/sentence-transformers/msmarco-hard-negatives}}. One could argue that this set of negatives may hurt the final performance on MS MARCO. However, it has been shown in \cite{splade_sigir22} that those negatives were delivering high performance for SPLADE. Using the model's own negatives improved mostly the performance on the BEIR benchmark but was not the best option for MS MARCO evaluation sets. Thus, the 
\textit{msmarco-hard-negatives} set was the best available set, and hence is our rationale for using it to compare models.

\subsection{Models}
We use  BERT-base-uncased as the PLM backbone. To see the impact of the datasets used during middle-training, we also added simple 'middle-trained' baselines: we middle-train BERT using the MS MARCO dataset with the normal MLM task (8 epochs as for RetroMAE, using the default parameters of the run\_mlm Hugging Face script.
We name this model \emph{BERT-MS}. Another simple baseline to better condition the CLS representation is to use the CLS token to predict the $L2$ normalized frequency of the input tokens similar to the term frequency regression in \cite{losses_pretraining_2020} and just adding this loss to the  MLM loss. We name this model \emph{BERT-MS-CLS}.
We limit our benchmark to the two main approaches for first-stage rankers: the dense and the sparse ones. The dense model simply computes the dot-product between the CLS tokens of both query and document to compute scores \cite{Hofstaetter2021_tasb_dense_retrieval,lin-etal-2021-batch}.  For the sparse approach, we use SPLADE \cite{formal_splade_2021,lassance2022efficiency}, with an L1 regularisation for the queries and a FLOPS one for documents\footnote{cf \url{https://github.com/naver/splade/tree/main/conf/efficient_splade}}.

\begin{table*}[ht]
\centering
\caption{
Evaluation for the no distillation, 1 negative setting. 
}
\resizebox{0.9\textwidth}{!}{
\begin{tabular}{clcccccccc}
\toprule
\multirow{3}{*}{\#}& \multirow{3}{*}{\textbf{Model}}   & \multicolumn{4}{c}{\textbf{DENSE}} &  \multicolumn{4}{c}{\textbf{SPLADE}}   \\ 
 \cmidrule(lr){3-6}  \cmidrule(lr){7-10} 
\textbf{}
& \textbf{}
&\multicolumn{2}{c}{\textbf{MSMARCO}}
&\multicolumn{1}{c}{\textbf{TREC19}}
&\multicolumn{1}{c}{\textbf{TREC20}}
&\multicolumn{2}{c}{\textbf{MSMARCO}}
&\multicolumn{1}{c}{\textbf{TREC19}}
&\multicolumn{1}{c}{\textbf{TREC20}}
\\
 \cmidrule(lr){3-4}  \cmidrule(lr){7-8} 
\textbf{}
& \textbf{}
& \textbf{MRR@10}
& \textbf{R@1K}
& \textbf{NDCG@10}
& \textbf{NDCG@10}
& \textbf{MRR@10}
& \textbf{R@1K}
& \textbf{NDCG@10}
& \textbf{NDCG@10} 
 \\

\midrule

a &
BERT &
31.9\hphantom{$^{bcdefg}$} & 
94.9\hphantom{$^{bcdefg}$} &
64.0\hphantom{$^{bcdefg}$} &
64.4\hphantom{$^{bcdefg}$} &
34.6\hphantom{$^{bcdefg}$} &  
96.5\hphantom{$^{bcdefg}$} &
69.6\hphantom{$^{bcdefg}$} &
67.6\hphantom{$^{bcdefg}$}  \\
\hdashline
b &
BERT MS &
33.4$^{a}$\hphantom{$^{cdefg}$} &
95.9$^{a}$\hphantom{$^{cdefg}$} &
67.0\hphantom{$^{acdefg}$} &
66.0\hphantom{$^{acdefg}$} &
35.8$^{a}$\hphantom{$^{cdefg}$} &
96.9\hphantom{$^{acdefg}$} &
68.3\hphantom{$^{acdefg}$} &
67.1\hphantom{$^{acdefg}$} \\

c &
BERT MS CLS &
33.7$^{a}$\hphantom{$^{bdefg}$} &
96.4$^{a}$\hphantom{$^{bdefg}$} &
66.9\hphantom{$^{abdefg}$} &
67.1\hphantom{$^{abdefg}$} &
35.8$^{a}$\hphantom{$^{bdefg}$} &
96.9\hphantom{$^{abdefg}$} &
68.2\hphantom{$^{abdefg}$} &
67.6\hphantom{$^{abdefg}$} \\

\hdashline
d &
coCondenser MS&
34.5$^{ab}$\hphantom{$^{cefg}$} &
97.2$^{abcg}$\hphantom{$^{ef}$} &
\textbf{68.8}\hphantom{$^{abcefg}$} &
67.1\hphantom{$^{abcefg}$} &
36.2$^{a}$\hphantom{$^{bcefg}$} &
97.3$^{a}$\hphantom{$^{bcefg}$} &
69.5\hphantom{$^{abcefg}$} &
67.1\hphantom{$^{abcefg}$} \\
e &
RetroMAE &
34.2$^{a}$\hphantom{$^{bcdfg}$} &
96.6$^{ab}$\hphantom{$^{cdfg}$} &
66.4\hphantom{$^{abcdfg}$} &
\textbf{67.7}\hphantom{$^{abcdfg}$} &
35.4\hphantom{$^{abcdfg}$} &
97.0\hphantom{$^{abcdfg}$} &
\textbf{70.2}\hphantom{$^{abcdfg}$} &
65.8\hphantom{$^{abcdfg}$} \\
 f & 
 RetroMAE MS &
 \textbf{35.9}$^{abcdeg}$\hphantom{} &
 \textbf{97.7}$^{abcdeg}$\hphantom{} &
 66.3\hphantom{$^{abcdeg}$} &
 67.7\hphantom{$^{abcdeg}$} &
36.3$^{a}$\hphantom{$^{bcdeg}$} &
\textbf{97.4}$^{ac}$\hphantom{$^{bdeg}$} &
69.6\hphantom{$^{abcdeg}$} &
66.3\hphantom{$^{abcdeg}$} \\
g &
LexMAE MS&
34.6$^{ab}$\hphantom{$^{cdef}$} &
96.6$^{ab}$\hphantom{$^{cdef}$} &
65.8\hphantom{$^{abcdef}$} &
65.7\hphantom{$^{abcdef}$} &
\textbf{36.8}$^{ae}$\hphantom{$^{bcdf}$} &
97.0\hphantom{$^{abcdef}$} &
70.2\hphantom{$^{abcdef}$} &
\textbf{68.4}\hphantom{$^{abcdef}$} \\
\bottomrule
\end{tabular}
}
\label{tab:densenodistil1neg}
\end{table*}

\begin{table}[ht]
\centering
\caption{
Evaluation for the no distillation, 32 negative setting
}
\resizebox{0.49\textwidth}{!}{
\begin{tabular}{clcccc}
\toprule
\multirow{2}{*}{\#}& \multirow{2}{*}{\textbf{Model}}   & \multicolumn{2}{c}{\textbf{Dense}} &  \multicolumn{2}{c}{\textbf{SPLADE}}   \\ 
\cmidrule(lr){3-4}  \cmidrule(lr){5-6} 
\textbf{}
& \textbf{}
& \footnotesize{ \textbf{MRR@10}}
&  \footnotesize{ \textbf{R@1K}}
&  \footnotesize{ \textbf{MRR@10}}
&  \footnotesize {\textbf{R@1K}} \\
\midrule
a &
BERT &
33.9\hphantom{$^{bcdefg}$} &
94.9\hphantom{$^{bcdefg}$} & 
36.3$^{ef}$\hphantom{$^{bcdg}$} &
96.2$^{cf}$\hphantom{$^{bdeg}$} \\
\hdashline
b &
BERT MS &
34.1\hphantom{$^{acdefg}$} &
95.9$^{a}$\hphantom{$^{cdefg}$} &
37.1$^{cef}$\hphantom{$^{adg}$} &
\textbf{97.2}$^{acef}$\hphantom{$^{dg}$} \\

c &
BERT MS CLS &
34.7\hphantom{$^{abdefg}$} &
95.7$^{a}$\hphantom{$^{bdefg}$} &
35.6$^{f}$\hphantom{$^{abdeg}$} &
95.5$^{f}$\hphantom{$^{abdeg}$} \\
\hdashline
d &
co-condenser-marco &
35.8$^{abc}$\hphantom{$^{efg}$} &
96.5$^{ac}$\hphantom{$^{befg}$} &
\textbf{38.0}$^{acef}$\hphantom{$^{bg}$} &
97.2$^{acef}$\hphantom{$^{bg}$} \\

e &
RetroMAE &
35.0$^{a}$\hphantom{$^{bcdfg}$} &
96.6$^{ac}$\hphantom{$^{bdfg}$} &
35.2$^{f}$\hphantom{$^{abcdg}$} &
95.7$^{f}$\hphantom{$^{abcdg}$} \\
f &
RetroMAE MS &
\textbf{35.9}$^{abc}$\hphantom{$^{deg}$} &
\textbf{97.0}$^{abcg}$\hphantom{$^{de}$} &
33.7\hphantom{$^{abcdeg}$} &
92.8\hphantom{$^{abcdeg}$} \\

g &
LexMAE &
35.9$^{abc}$\hphantom{$^{def}$} &
96.1$^{a}$\hphantom{$^{bcdef}$} &
37.4$^{acef}$\hphantom{$^{bd}$} &
97.1$^{acef}$\hphantom{$^{bd}$} \\

\bottomrule
\end{tabular}
}
\label{tab:nodistl32}
\end{table}

\section{Experiments}
We explain in this section how the finetuning steps are performed with all the MTLMs. The aim is to show how they behave with different datasets and different finetuning ``settings''.

We remind that we use the \textit{msmarco-hard-negatives} for all MS MARCO experiments, and our two main features are whether to use distillation or not and the number of hard negatives. 
First, using 1 negative and non-distillation. Then, 1 negative and distillation, and finally using 32 negatives and distillation. If these settings are not exhaustive, they cover the most frequent settings found in the literature. One goal of using various settings is to see how the different models behave, and if one setting favors some models.
For the TripClick dataset, we use the negatives provided by \cite{hofstaetter2022tripclick}, and just test the no distillation, 1 negative setting.

We have used open-source repositories to set up this benchmark: sentence Transformer\footnote{https://github.com/UKPLab/sentence-transformers}, Hugging Face and  SPLADE github\footnote{Code available at \url{https://github.com/naver/splade/tree/benchmarch-SIGIR23} } where the various configurations used in this benchmark are available.
For each setting, one single set of hyperparameters is used. A learning rate of $2\mathrm{e}{-5}$ is used for all experiments. For the settings with 1 negative, we simply use the default regularization values provided by the SPLADE GitHub (the main ones being a number of iterations of 150.000, a batch size of 128, and a max length of 256). For the distillation setting with 1 negative, we use the MSE loss. For the 32 negatives version, we have largely changed the values to optimize the memory usage: we use 5 epochs, a batch size per device of 170 (5 queries), a max length of 128, and the KL-Div loss as we found that the MSE loss was not the best with several negatives. 
We have found these settings effective for all our experiments, even if we acknowledge that they may be sub-optimal for some  models we tested.  Experiments were conducted using 4 V100 32Gb.

First, we evaluate this benchmark using the traditional dataset: MS MARCO (dev) \cite{nguyen2016ms} without titles as in the official setting (cf \cite{lassance2023tale}, TREC'19 and TREC'20. For zero-shot experiments, we use the 13 public sub-datasets of the BEIR dataset \cite{thakurbeir2021}. 
To test the finetuning of these models in a completely different domain, we use the Tripclick dataset (medical domain) \cite{rekabsaz2021fairnessir}.
For evaluation, we use the RANX python package~\cite{ranx}. We use MRR@10 and R@1K for the MS MARCO dataset and nDCG@10 for the TREC'19 and TREC'20 datasets. These values have been multiplied by 100. The BEIR evaluation corresponds to the means of the nDCG@10 over the 13 public datasets.
For all settings but BEIR, we perform a paired Student's t-test (alpha=0.05). We adjust the desired alpha level by the number of comparisons (Bonferroni correction \cite{Blan1995170}). Superscripts denote significant differences. The best results are in bold.

\subsection{MS MARCO}
For each setting, we will show its result in a table, comparing the various models with both approaches: dense (left) and sparse (right). In all settings, the sparse approach beats the dense one.  We separate the three model classes: (a,b): off-the-shelf models. (c,d) basic MTLMs, (e,f,g): the three MTLMs found in the literature. 
For all our sets of experiments, no statistical difference is detected for the TREC'19 ad TREC'20 datasets, the margin being too small for the limited set of queries. For this reason, we show them only  on the first table.

\paragraph{ Table~\ref{tab:densenodistil1neg}: no distillation, 1  negative} 
First, as expected, the BERT baseline models perform poorly compared to the middle-trained ones, but we see with the basic MTLMs (b,c) the impact of using MS MARCO during this middle-training step gives some boost($>$+1.4 point).
On the dense side, RetroMAE MS beats all the other models, while on the sparse side, LexMAE achieves the best score, but is on par with most of the other models.

For this setting, the importance of middle-training on the downstream collection is clearly shown by comparing the versions of the models which have/haven't seen MS MARCO. For instance, between BERT and BERT MS/CLS(a,b-c), but also between RetroMAE and its MS version (e,f). 
Finally, the difference between RetroMAE and its MS version is noticeable, demonstrating the importance of middle training on the retrieval collection.

\paragraph{ Table~\ref{tab:nodistl32}: no distillation, 32  negatives}
In this setting, the use of several negatives introduces some unexpected behavior for some models. For the dense approach, almost all models improve their results in terms of MRR but recall at 1000 stays at the same level or even decreases. For sparse models, results can improve  or decrease depending on the models. In this case, a specific finetuning for each model seems to be required here.

\begin{table}[ht]
\centering
\caption{ 
Evaluation for the distillation, 1 negative setting. 
}

\resizebox{0.49 \textwidth}{!}{
\begin{tabular}{clcccccc}
\toprule
\multirow{2}{*}{\#}& \multirow{2}{*}{\textbf{Model}}   & \multicolumn{3}{c}{\textbf{Dense}} &  \multicolumn{3}{c}{\textbf{SPLADE}}   \\ 
\cmidrule(lr){3-5}  \cmidrule(lr){6-8} 
\textbf{} 
& \textbf{}
& \footnotesize{\textbf{MRR@10}}
&  \footnotesize{\textbf{R@1k}}
& \footnotesize {\textbf{BEIR}}
& \footnotesize{ \textbf{MRR@10}}
&  \footnotesize {\textbf{R@1k}}
&  \footnotesize {\textbf{BEIR}} \\
\midrule
a &
BERT &
35.3\hphantom{$^{bcdefg}$} &
97.3\hphantom{$^{bcdefg}$}  &
 44.0  &
37.8\hphantom{$^{bcdefg}$} &
97.8\hphantom{$^{bcdefg}$} &
 49.6  \\
 \hdashline
b &
BERT MS &
36.2\hphantom{$^{acdefg}$} &
97.9$^{a}$\hphantom{$^{cdefg}$}   &
 44.2  &
38.3\hphantom{$^{acdefg}$} &
98.0\hphantom{$^{acdefg}$} &
 49.6  \\

c &
BERT MS CLS &
36.5$^{a}$\hphantom{$^{bdefg}$} &
98.0$^{a}$\hphantom{$^{bdefg}$}  &
44.2   &
38.4\hphantom{$^{abdefg}$} &
98.1$^{a}$\hphantom{$^{bdefg}$}  &
 49.6  \\
 \hdashline
d &
coCondenser MS &
36.9$^{a}$\hphantom{$^{bcefg}$} &
98.0$^{a}$\hphantom{$^{bcefg}$}  & 
46.6   &
\textbf{38.6}$^{a}$\hphantom{$^{bcefg}$} &
98.2$^{ae}$\hphantom{$^{bcfg}$}   &
 \textbf{50.0}  \\
e &
RetroMAE &
36.7$^{a}$\hphantom{$^{bcdfg}$} &
98.0$^{a}$\hphantom{$^{bcdfg}$}  &
46.8  &
38.1\hphantom{$^{abcdfg}$} &
98.0\hphantom{$^{abcdfg}$}  &
 49.3  \\
f& RetroMAE MS & 
\textbf{37.7}$^{abce}$\hphantom{$^{dg}$} &
\textbf{98.3}$^{aceg}$\hphantom{$^{bd}$} & 
46.2&
38.5\hphantom{$^{abcdeg}$} &
\textbf{98.3}$^{ae}$\hphantom{$^{bcdg}$} & 
49.8   \\  
g &
LexMAE MS&
36.8$^{a}$\hphantom{$^{bcdef}$} &
97.9$^{a}$\hphantom{$^{bcdef}$}  & 
 44.3  &
38.4\hphantom{$^{abcdef}$} &
98.1$^{a}$\hphantom{$^{bcdef}$}  &
 49.6   \\
\bottomrule
\end{tabular}
}
\label{tab:densdistl1neg}
\end{table}

\paragraph{Table~\ref{tab:densdistl1neg}: distillation, 1  negative} Thanks to distillation, the models' effectiveness increases by almost 3 points.  RetroMAE MS keeps its advantages but is now on par with coCondenser and LexMAE.
For SPLADE, the basic MTLM (d) is already competitive, and LexMAE's advantage shown in Table~\ref{tab:densenodistil1neg} disappears to the benefit of coCondenser. 


\begin{table}[ht]
\centering
\caption{ Evaluation for the distillation, 32 negative setting. 
}
\resizebox{0.49\textwidth}{!}{
\begin{tabular}{clccccc}
\toprule
\multirow{2}{*}{\#}& \multirow{2}{*}{\textbf{Model}}   & \multicolumn{2}{c}{\textbf{Dense}} &  \multicolumn{2}{c}{\textbf{SPLADE}}   \\ 
\cmidrule(lr){3-4}  \cmidrule(lr){5-6} 
\textbf{}
& \textbf{}
& \footnotesize{ \textbf{MRR@10}}
&  \footnotesize{ \textbf{R@1K}}
&  \footnotesize{ \textbf{MRR@10}}
&  \footnotesize {\textbf{R@1K}} 
&  \footnotesize {\textbf{BEIR}}\\
\midrule

a &
BERT &
37.4\hphantom{$^{bcdefg}$} &
97.0\hphantom{$^{bcdefg}$} &

39.1\hphantom{$^{bcdefg}$} &
97.7\hphantom{$^{bcdefg}$} & 
45.7 \\
\hdashline
b &
BERT MS &
37.7\hphantom{$^{acdefg}$} &
97.4$^{a}$\hphantom{$^{cdefg}$} &
39.4\hphantom{$^{acdefg}$} &
97.9\hphantom{$^{acdefg}$}  &
45.9 \\
c &
BERT MS CLS &
38.0\hphantom{$^{abdefg}$} &
97.5$^{a}$\hphantom{$^{bdefg}$}  &
39.7\hphantom{$^{abdefg}$} &
98.0\hphantom{$^{abdefg}$} &
45.23\\
\hdashline
d &
coCondenser MS &
38.1\hphantom{$^{abcefg}$} &
98.1$^{abcg}$\hphantom{$^{ef}$}&
39.6\hphantom{$^{abcefg}$} &
97.9\hphantom{$^{abcefg}$} &
\textbf{46.4}\\
e &
RetroMAE &
38.6$^{ab}$\hphantom{$^{cdfg}$} &
98.0$^{abcg}$\hphantom{$^{df}$} &
39.3\hphantom{$^{abcdfg}$} &
97.9\hphantom{$^{abcdfg}$} &
45.8\\
f& RetroMAE MS  &
\textbf{39.2}$^{abcdg}$\hphantom{$^{e}$} &
\textbf{98.3}$^{abcg}$\hphantom{$^{de}$} &
39.5\hphantom{$^{abcdeg}$} &
\textbf{98.4}$^{abcde}$\hphantom{$^{g}$} &
  40.6
 \\
g &
LexMAE MS&
38.3$^{a}$\hphantom{$^{bcdef}$} &
97.5$^{a}$\hphantom{$^{bcdef}$} &
\textbf{39.8}$^{a}$\hphantom{$^{bcdef}$} &
98.1\hphantom{$^{abcdef}$} &
45.5\\
\bottomrule
\end{tabular}
    }
\label{tab:densedsitl32}
\end{table}

\paragraph{Table~\ref{tab:densedsitl32}: distillation, 32  negatives} For this most effective setting, the differences between MTLMs continue to shrink, almost disappearing for the sparse approach: even  BERT performs honorably, and it is difficult to select a winner.
In terms of MRR, RetroMAE MS is the very best dense configuration and reaches the level of the sparse models, which shows that middle-training has a positive effect.  Focusing on recall (at 1000), middle-trained dense models outperform the basic MTLM ones, and the sparse models, except the winner RetroMAE MS, all performs similarly. 


\paragraph{Zero-Shot Evaluation on BEIR: Table~\ref{tab:densdistl1neg} and \ref{tab:densedsitl32} }  
As we can see on the BEIR column for table \ref{tab:densdistl1neg} , it is difficult to find the best configuration(s): for the sparse setting, the off-the-shelf BERT  already performs very well (49.6), and coCondenser is slightly better (50.0), showing that middle-training does not help much in the zero-shot setting. Table~\ref{tab:densedsitl32} also shows results for the BEIR dataset (SPLADE only) with 32 negatives, which are quite disappointing. This setting  considers a max length of 128, due to the batch size, which impacts a lot the performances on BEIR. By testing with 256 tokens at retrieval time, results improve and reach the ones of the simpler setting using 1 negative (results not shown in this benchmark). But for our purpose of comparing MTLMs, we see that the most enhanced MTLMs do not provide better generalisation on BEIR.

Generally, we can note that {\it i)} our middle-trained baselines (b,c) are competitive in almost all the previous settings, and  {\it ii)} sparse MTLMs work well for dense and dense MTLMs work well for sparse. We hypothesize that this is due to the implicit LM adaptation performed during middle training.

\paragraph{On Reproducibility:}
an expert reader may have spotted that the reported results for MTLMs differ from their original papers. The first explanation is that some models, after inspection of their GitHub repository, do rely on a modified MS MARCO that uses the title information, which is different from the official MS MARCO. Please consider that this issue has actually been noticed by several people\footnote{\url{https://twitter.com/macavaney/status/1627373586833981441}  \url{https://github.com/texttron/tevatron/issues/34}} 
and therefore motivate the need for a fair benchmark \cite{lassance2023tale}. A second explanation could be a different hyperparameter tuning, cross-encoder teachers, and set of hard negatives, which we have tried to mitigate by using an ensemble of negatives and the same teacher.

\begin{table}[ht]
\centering
\caption{ 50K/100K queries; distillation, 32 neg. setting (MRR@10)
}
\resizebox{0.48\textwidth}{!}{
\begin{tabular}{clcccc}
\toprule
\multirow{2}{*}{\#}& \multirow{2}{*}{\textbf{Model}}   & \multicolumn{2}{c}{\textbf{Dense}} &  \multicolumn{2}{c}{\textbf{SPLADE}}   \\ 
\cmidrule(lr){3-4}  \cmidrule(lr){5-6} 
\textbf{}
& \textbf{}
& \textbf{50k}
& \textbf{100k}
& \textbf{50k}
& \textbf{100K} 
\\
\midrule
a &
BERT &
33.2\hphantom{$^{bcdefg}$}  &
34.6\hphantom{$^{bcdefg}$}  &
36.2\hphantom{$^{bcdefg}$}& 
38.1\hphantom{$^{bcdefg}$} 
\\
\hdashline
b &
BERT MS &
35.3$^{a}$\hphantom{$^{cdefg}$}  &
36.5$^{a}$\hphantom{$^{cdefg}$} &
37.4$^{a}$\hphantom{$^{cdefg}$}  &
38.4\hphantom{$^{acdefg}$}  
\\
c &
BERT MS CLS &
35.4$^{a}$\hphantom{$^{bdefg}$}& 
36.4$^{a}$\hphantom{$^{bdefg}$} &
37.7$^{ae}$\hphantom{$^{bdfg}$} & 
38.7\hphantom{$^{abdefg}$}  
\\
\hdashline
d &
coCondenser MS &
36.7$^{abc}$\hphantom{$^{efg}$}&
37.2$^{ac}$\hphantom{$^{befg}$}  &
38.0$^{ae}$\hphantom{$^{bcfg}$}&
38.9$^{ae}$\hphantom{$^{bcfg}$} 
\\
e &
RetroMAE &
36.2$^{a}$\hphantom{$^{bcdfg}$} &
\textbf{37.4}$^{abc}$\hphantom{$^{dfg}$} &
36.9$^{a}$\hphantom{$^{bcdfg}$} &
38.2\hphantom{$^{abcdfg}$}  

\\
f & RetroMAE MS &
\textbf{37.0}$^{abcg}$\hphantom{$^{de}$} &
37.2$^{a}$\hphantom{$^{bcdeg}$}  &
37.8$^{ae}$\hphantom{$^{bcdg}$} &
39.0$^{ae}$\hphantom{$^{bcdg}$}   
 \\
g &
LexMAE &
36.0$^{a}$\hphantom{$^{bcdef}$}&
37.2$^{a}$\hphantom{$^{bcdef}$} &
\textbf{38.3}$^{abe}$\hphantom{$^{cdf}$}&
\textbf{39.3}$^{abe}$\hphantom{$^{cdf}$} 

\\
\bottomrule
\end{tabular}
}
\label{tab:Xqueries}
\end{table}

\begin{table*}[ht]
\centering
\caption{
50K/100K queries; distillation, 32 neg. setting (sparse only)
}
\resizebox{0.8\textwidth}{!}{
\begin{tabular}{llcccccccc}
\toprule
\multirow{2}{*}{\#}& \multirow{2}{*}{\textbf{Model}}   & \multicolumn{4}{c}{\textbf{50K}} &  \multicolumn{4}{c}{\textbf{100K}}   \\ 
\cmidrule(lr){3-6}  \cmidrule(lr){7-10} 
& \textbf{}
& \textbf{MRR@10}
& \textbf{Recall@100}
& \textbf{Recall@1K}
& \textbf{BEIR}
& \textbf{MRR@10}
& \textbf{Recall@100}
& \textbf{Recall@1K} 
& \textbf{BEIR}\\
\midrule
a &
BERT &
33.2\hphantom{$^{bcdefg}$} &
84.4\hphantom{$^{bcdefg}$} &
94.7\hphantom{$^{bcdefg}$} &
39.74  &
38.1\hphantom{$^{bcdefg}$} &
88.9\hphantom{$^{bcdefg}$} &
97.3\hphantom{$^{bcdefg}$} &
44.21\\
\hdashline
b &
BERT MS &
37.4$^{a}$\hphantom{$^{cdefg}$} &
88.8$^{a}$\hphantom{$^{cdefg}$} &
97.4$^{a}$\hphantom{$^{cdefg}$} &
46.51  &
38.4\hphantom{$^{acdefg}$} &
89.7$^{a}$\hphantom{$^{cdefg}$} &
97.5\hphantom{$^{acdefg}$}  &
46.80\\
c &
BERT MS CLS &
37.7$^{ae}$\hphantom{$^{bdfg}$} &
\textbf{89.4}$^{ae}$\hphantom{$^{bdfg}$} &
\textbf{97.6}$^{ae}$\hphantom{$^{bdfg}$} &
45.71 &
38.7\hphantom{$^{abdefg}$} &
\textbf{90.3}$^{abe}$\hphantom{$^{dfg}$} &
\textbf{97.8}$^{ae}$\hphantom{$^{bdfg}$} &
46.82\\
\hdashline
d &
coCondenser MS&
38.0$^{ae}$\hphantom{$^{bcfg}$} &
89.3$^{ae}$\hphantom{$^{bcfg}$} &
97.5$^{a}$\hphantom{$^{bcefg}$} &
\textbf{47.23} &
38.9$^{a}$\hphantom{$^{bcefg}$} &
89.8$^{a}$\hphantom{$^{bcefg}$} &
97.6\hphantom{$^{abcefg}$} &
\textbf{47.72}\\
e &
RetroMAE &
36.9$^{a}$\hphantom{$^{bcdfg}$} &
88.2$^{a}$\hphantom{$^{bcdfg}$} &
97.1$^{a}$\hphantom{$^{bcdfg}$} &
46.30 &
38.2\hphantom{$^{abcdfg}$} &
89.2\hphantom{$^{abcdfg}$} &
97.4\hphantom{$^{abcdfg}$} &
46.58\\
f &
RetroMAE-MS &
37.8$^{ae}$\hphantom{$^{bcdg}$} &
89.3$^{ae}$\hphantom{$^{bcdg}$} &
97.4$^{a}$\hphantom{$^{bcdeg}$} &
46.41 &
39.0$^{ae}$\hphantom{$^{bcdg}$} &
90.2$^{ae}$\hphantom{$^{bcdg}$} &
97.8$^{ae}$\hphantom{$^{bcdg}$} &
47.14\\
g &
LexMAE &
\textbf{38.3}$^{abe}$\hphantom{$^{cdf}$} &
89.4$^{ae}$\hphantom{$^{bcdf}$} &
97.5$^{a}$\hphantom{$^{bcdef}$} &
46.75 &
\textbf{39.3}$^{abe}$\hphantom{$^{cdf}$} &
90.0$^{ae}$\hphantom{$^{bcdf}$} &
97.6\hphantom{$^{abcdef}$} &
47.28\\
\bottomrule
\end{tabular}
}
\label{tab:sparse_queries_recall}
\end{table*}



\subsection{Finetuning with fewer Queries}
An evaluation where middle-trained models could be more effective and relevant is a more realistic one than the MS MARCO setting, where half a million queries are available. We then reduce this amount down to a random 50k and 100k query subsets\footnote{available at \url{https://github.com/naver/splade/tree/benchmarch-SIGIR23}. } and then finetune the models. We see in Table~\ref{tab:Xqueries} that the observations are very similar to the "full queries" setting: for the dense approach, coCondenser and RetroMAE MS perform significantly better than the 3 baselines, the best being RetroMAE MS. For the sparse approach, LexMAE is the winner, but there is no statistical difference with some other models, among them the BERT-MS-CLS baseline.
We can note that the gap between the best models and the basic ones increases when using less queries. 
Table~\ref{tab:sparse_queries_recall} shows more detailed evaluation for the SPLADE models: we can see that the various MTLMs reach  similar values in terms of recall, the BERT MS CLS model reaching the best scores, but without statistical difference (except with BERT).


\begin{table}[ht]
\centering
\caption{
Tripclick Evaluation (nDCG@10) for the no distillation, 1 negative setting (dense only). Middle-trained models do not help on a different collection.
}
\resizebox{0.48\textwidth}{!}{
\begin{tabular}{clcccc}
\toprule
& \textbf{Model}
& \textbf{Head$_{dctr}$}
& \textbf{Head}
& \textbf{Torso}
& \textbf{Tail} \\
\midrule
a &
BERT &
25.3\hphantom{$^{bcdefghi}$} &
35.3\hphantom{$^{bcdefghi}$} &
29.4\hphantom{$^{bcdefghi}$} &
28.8\hphantom{$^{bcdefghi}$} \\
\hdashline
b &
PubMedBert &
27.1$^{ai}$\hphantom{$^{cdefgh}$} &
37.6$^{ai}$\hphantom{$^{cdefgh}$} &
29.9\hphantom{$^{acdefghi}$} &
30.8$^{ei}$\hphantom{$^{acdfgh}$} \\
c &
Scibert &
27.8$^{aefi}$\hphantom{$^{bdgh}$} &
38.1$^{aefi}$\hphantom{$^{bdgh}$} &
29.2\hphantom{$^{abdefghi}$} &
30.3$^{i}$\hphantom{$^{abdefgh}$} \\
d &
BERT MLM 12L &
\textbf{28.0}$^{aefgi}$\hphantom{$^{bch}$} &
\textbf{38.6}$^{aefgi}$\hphantom{$^{bch}$} &
30.2\hphantom{$^{abcefghi}$} &
30.4$^{i}$\hphantom{$^{abcefgh}$} \\
\hdashline
e &
BERT MS &
26.1\hphantom{$^{abcdfghi}$} &
36.5\hphantom{$^{abcdfghi}$} &
30.3$^{i}$\hphantom{$^{abcdfgh}$} &
28.3\hphantom{$^{abcdfghi}$} \\
f &
BERT MS CLS &
26.2\hphantom{$^{abcdeghi}$} &
36.3\hphantom{$^{abcdeghi}$} &
30.0$^{i}$\hphantom{$^{abcdegh}$} &
28.9\hphantom{$^{abcdeghi}$} \\
g &
coCondenser MS &
26.6$^{a}$\hphantom{$^{bcdefhi}$} &
36.6\hphantom{$^{abcdefhi}$} &
30.1$^{i}$\hphantom{$^{abcdefh}$} &
29.5\hphantom{$^{abcdefhi}$} \\
h &
RetroMAE &
27.5$^{aei}$\hphantom{$^{bcdfg}$} &
37.9$^{afi}$\hphantom{$^{bcdeg}$} &
\textbf{31.0}$^{i}$\hphantom{$^{abcdefg}$} &
\textbf{31.2}$^{aefi}$\hphantom{$^{bcdg}$} \\
i &
LexMAE MS&
25.6\hphantom{$^{abcdefgh}$} &
35.3\hphantom{$^{abcdefgh}$} &
28.3\hphantom{$^{abcdefgh}$} &
27.5\hphantom{$^{abcdefgh}$} \\
\bottomrule
\end{tabular}
}
\label{tab:tripclick}
\end{table}

\subsection{TripClick}
We finally discuss how these MTLMs, using MS MARCO as training data, behave in an out-of-domain setting by finetuning them with the TripClick dataset \cite{rekabsaz2021fairnessir}, from the medical domain. This dataset interestingly divides the queries set by frequency (Head, Torso, Tail).
We compare our MTLMs with some results obtained in \cite{lassance_experimental_2023} where they "pre-train from scratch"  different Transformer models. Due to lack of space, only the dense models are shown in Table~\ref{tab:tripclick}
Rows (b,c,d) are from \cite{pretraining_scratch_arxiv23}, (d) being trained from scratch using TripClick. We see that our models middle-trained on MS MARCO (e,f) overall underperform (as expected) compared to (b,c,d), but nevertheless, improve over BERT (a). The good performance of RetroMAE, trained only on WikiBook data,  is noticeable, especially for low-frequency queries. We see that LexMAE performs poorly, but this is explainable due to its characteristics: trained with MS MARCO and for a sparse configuration.
In this experiment, the MTLMs are indeed pre-trained to handle retrieval better, but their LM part is not well adapted to the collection. Except for RetroMAE, trained on Wikipedia, these models perform worst compared to in-domain language models, showing again the importance of middle training on the retrieval collection.

\section{DISCUSSION}
We now discuss our results to highlight the main findings.
First, our results on SPLADE reveals that there is almost no statistical difference between those methods: the more effective the finetuning procedure is, the less difference there is between those models as illustrated in Figure \ref{fig:splade}.
For the dense RetroMAE-MS is always better the baseline and other middle training methods. However, if RetroMAE is not trained on the retrieval collection, it looses its competitive advantage as shows in Table \ref{tab:densedsitl32} and on the TripClick experiments.
Surprisingly, the dense pretraining works for sparse models and the sparse middle training works for dense models as well. One explication could be that adapting the language model is actually a key component of the middle-training step, by the adjusting the co-occurrence statistics for the retrieval collection.

\section{CONCLUSION}
In this paper, we have benchmarked, in the same conditions, several middle-trained models with various finetuning settings and datasets. 
In these experiments, results for the sparse and dense approaches are divergent.
For SPLADE, all the models that have used the downstream collection are very close, even if LexMAE usually gets the best results. It may indicate that the proper way to middle-train a sparse model needs to be further investigated.
For dense, RetroMAE MS stands out in all MS MARCO experiments, 
Some of our experiments ( with fewer queries) show that in some settings, middle-trained models perform better than the middle-trained baselines. 

We find it difficult to assess the main factor of the middle-trained models: the benefits of their specific architecture and loss. Strangely middle-trained models designed for dense approaches (through CLS token), such as coCondenser and RetroMAE, have a positive effect also for the sparse setting. Additionally, we noted (not reported in this article) that with the most effective setting (using several hard negatives), the middle-trained models were usually easier to fine-tune with SPLADE.

The key point of this benchmark is the importance of the collection used during middle-training, which has a strong impact, as shown by the differences between RetroMAE and RetroMAE MS, but also shown with our middle-trained baselines: by just middle-training it with the target corpus, a BERT model becomes almost competitive. Adapting the language model is then a key component of the middle-training step, at least as important as the added IR-related tasks or the architecture modification. This is in line with work which shows that pretraining from scratch a model with the target corpus is also a effective approach \cite{lassance_experimental_2023}.

In general, a clearer setting is required in order to allow for a fair and more precise comparison between these models.
To conclude, we recommend releasing the set of hard negatives, and the cross-encoder teachers to ease reproducibility and to study the benefit of middle training under few shot conditions or limited finetuning data.





\bibliographystyle{ACM-Reference-Format}
\balance{}
\bibliography{sample-base}

\end{document}